\documentclass[fleqn,usenatbib]{mnras}

\usepackage{graphicx}	% Including figure files
\usepackage{amsmath}	% Advanced maths commands
\usepackage{amssymb}	% Extra maths symbols
\usepackage{rotating}
\usepackage{pdflscape}
\usepackage{longtable}
\usepackage{afterpage}
\usepackage[T1]{fontenc}
\usepackage{booktabs}
\usepackage{xcolor}
\usepackage{eso-pic}% http://ctan.org/pkg/eso-pic

\AddToShipoutPictureBG*{%
  \AtPageUpperLeft{%
    \hspace{0.75\paperwidth}%
    \raisebox{-3.5\baselineskip}{%
      \makebox[0pt][l]{\textnormal{DES-2023-0762}}
}}}%

\AddToShipoutPictureBG*{%
  \AtPageUpperLeft{%
    \hspace{0.75\paperwidth}%
    \raisebox{-4.5\baselineskip}{%
      \makebox[0pt][l]{\textnormal{FERMILAB-PUB-23-372}}
}}}%

\newcommand{\orcid}[1]{\href{https://orcid.org/#1}{\includesvg[width=10pt]{orcid}}}%%%%%%%%%%%%%%%%%%%%%%%%%%%%%%%%%%%%%%%%%%%%%%%%

\title[Taxonomical information from DES data]
      {Main belt asteroids taxonomical information from Dark Energy Survey data}
\author[DES Collaboration]{
\parbox{\textwidth}{
\Large
V.~Carruba,$^{1,2}$
J.~I.~B.~Camargo,$^{3,2}$
S.~Aljbaae,$^{4}$
F.~S.~Ferreira,$^{3,2}$
E.~Lin,$^{5}$
V.~Figueiredo-Peixoto,$^{3,6,2}$
M.~V.~Banda-Huarca,$^{3,2}$
A.~Pieres,$^{2}$
R.~C.~Boufleur,$^{3,2}$
L.~N.~da Costa,$^{2}$
T.~M.~C.~Abbott,$^{7}$
M.~Aguena,$^{2}$
Sahar~S.~Allam,$^{8}$
O.~Alves,$^{5}$
P.~H.~Bernardinelli,$^{9}$
E.~Bertin,$^{10,11}$
D.~Brooks,$^{12}$
A.~Carnero~Rosell,$^{13,2,14}$
J.~Carretero,$^{15}$
M.~E.~S.~Pereira,$^{16}$
T.~M.~Davis,$^{17}$
J.~De~Vicente,$^{18}$
S.~Desai,$^{19}$
P.~Doel,$^{12}$
I.~Ferrero,$^{20}$
D.~Friedel,$^{21}$
J.~Frieman,$^{8,22}$
J.~Garc\'ia-Bellido,$^{23}$
M.~Gatti,$^{24}$
G.~Giannini,$^{15}$
D.~Gruen,$^{25}$
R.~A.~Gruendl,$^{21,26}$
K.~Herner,$^{8}$
S.~R.~Hinton,$^{17}$
D.~L.~Hollowood,$^{27}$
D.~J.~James,$^{28}$
S.~Kent,$^{8,22}$
K.~Kuehn,$^{29,30}$
O.~Lahav,$^{12}$
J.~L.~Marshall,$^{31}$
J. Mena-Fern{\'a}ndez,$^{18}$
R.~Miquel,$^{32,15}$
A.~Palmese,$^{33}$
A.~A.~Plazas~Malag\'on,$^{34,35}$
M.~Rodr\'iguez-Monroy,$^{36}$
E.~Sanchez,$^{18}$
B.~Santiago,$^{37,2}$
M.~Schubnell,$^{5}$
M.~Smith,$^{38}$
E.~Suchyta,$^{39}$
M.~E.~C.~Swanson,$^{40}$
G.~Tarle,$^{5}$
A.~R.~Walker,$^{7}$
N.~Weaverdyck,$^{5,41}$
and P.~Wiseman$^{38}$
\begin{center} (DES Collaboration) \end{center}
}
\vspace{0.4cm}
\\
\parbox{\textwidth}{
$^{1}$ S\~{a}o Paulo State University (UNESP), School of Natural Sciences and Engineering\\
$^{2}$ Laborat\'orio Interinstitucional de e-Astronomia - LIneA, Brazil\\
$^{3}$ Observat\'orio Nacional, Rua Gal. Jos\'e Cristino 77, Rio de Janeiro, RJ - 20921-400, Brazil\\
$^{4}$ National Space Research Institute (INPE), Division of Space Mechanics and Control, C.P. 515, 12227-310, S\~{a}o Jos\'e dos Campos, SP, Brazil\\
$^{5}$ Department of Physics, University of Michigan, Ann Arbor, MI 48109, USA\\
$^{6}$ IGeo - Universidade Federal do Rio de Janeiro, Av. Athos da Silveira Ramos 274, Rio de Janeiro, RJ - 21941-916, Brazil\\
$^{7}$ Cerro Tololo Inter-American Observatory, NSF's National Optical-Infrared Astronomy Research Laboratory, Casilla 603, La Serena, Chile\\
$^{8}$ Fermi National Accelerator Laboratory, P. O. Box 500, Batavia, IL 60510, USA\\
$^{9}$ Astronomy Department, University of Washington, Box 351580, Seattle, WA 98195, USA\\
$^{10}$ CNRS, UMR 7095, Institut d'Astrophysique de Paris, F-75014, Paris, France\\
$^{11}$ Sorbonne Universit\'es, UPMC Univ Paris 06, UMR 7095, Institut d'Astrophysique de Paris, F-75014, Paris, France\\
$^{12}$ Department of Physics \& Astronomy, University College London, Gower Street, London, WC1E 6BT, UK\\
$^{13}$ Instituto de Astrofisica de Canarias, E-38205 La Laguna, Tenerife, Spain\\
$^{14}$ Universidad de La Laguna, Dpto. AstrofÃ­sica, E-38206 La Laguna, Tenerife, Spain\\
$^{15}$ Institut de F\'{\i}sica d'Altes Energies (IFAE), The Barcelona Institute of Science and Technology, Campus UAB, 08193 Bellaterra (Barcelona) Spain\\
$^{16}$ Hamburger Sternwarte, Universit\"{a}t Hamburg, Gojenbergsweg 112, 21029 Hamburg, Germany\\
$^{17}$ School of Mathematics and Physics, University of Queensland,  Brisbane, QLD 4072, Australia\\
$^{18}$ Centro de Investigaciones Energ\'eticas, Medioambientales y Tecnol\'ogicas (CIEMAT), Madrid, Spain\\
$^{19}$ Department of Physics, IIT Hyderabad, Kandi, Telangana 502285, India\\
$^{20}$ Institute of Theoretical Astrophysics, University of Oslo. P.O. Box 1029 Blindern, NO-0315 Oslo, Norway\\
$^{21}$ Center for Astrophysical Surveys, National Center for Supercomputing Applications, 1205 West Clark St., Urbana, IL 61801, USA\\
$^{22}$ Kavli Institute for Cosmological Physics, University of Chicago, Chicago, IL 60637, USA\\
$^{23}$ Instituto de Fisica Teorica UAM/CSIC, Universidad Autonoma de Madrid, 28049 Madrid, Spain\\
$^{24}$ Department of Physics and Astronomy, University of Pennsylvania, Philadelphia, PA 19104, USA\\
$^{25}$ University Observatory, Faculty of Physics, Ludwig-Maximilians-Universit\"at, Scheinerstr. 1, 81679 Munich, Germany\\
$^{26}$ Department of Astronomy, University of Illinois at Urbana-Champaign, 1002 W. Green Street, Urbana, IL 61801, USA\\
$^{27}$ Santa Cruz Institute for Particle Physics, Santa Cruz, CA 95064, USA\\
$^{28}$ Center for Astrophysics $\vert$ Harvard \& Smithsonian, 60 Garden Street, Cambridge, MA 02138, USA\\
$^{29}$ Australian Astronomical Optics, Macquarie University, North Ryde, NSW 2113, Australia\\
$^{30}$ Lowell Observatory, 1400 Mars Hill Rd, Flagstaff, AZ 86001, USA\\
$^{31}$ George P. and Cynthia Woods Mitchell Institute for Fundamental Physics and Astronomy, and Department of Physics and Astronomy, Texas A\&M University, College Station, TX 77843,  USA\\
$^{32}$ Instituci\'o Catalana de Recerca i Estudis Avan\c{c}ats, E-08010 Barcelona, Spain\\
$^{33}$ Department of Physics, Carnegie Mellon University, Pittsburgh, Pennsylvania 15312, USA\\
$^{34}$ Kavli Institute for Particle Astrophysics \& Cosmology, P. O. Box 2450, Stanford University, Stanford, CA 94305, USA\\
$^{35}$ SLAC National Accelerator Laboratory, Menlo Park, CA 94025, USA\\
$^{36}$ Laboratoire de Physique des 2 Infinis Ir\`ene Joliot-Curie,CNRS Universit\'e Paris-Saclay, B\^at. 100, Facult\'e des sciences, F-91405 Orsay Cedex, France\\
$^{37}$ Instituto de F\'\i sica, UFRGS, Caixa Postal 15051, Porto Alegre, RS - 91501-970, Brazil\\
$^{38}$ School of Physics and Astronomy, University of Southampton,  Southampton, SO17 1BJ, UK\\
$^{39}$ Computer Science and Mathematics Division, Oak Ridge National Laboratory, Oak Ridge, TN 37831\\
$^{40}$ Independent Researcher\\
$^{41}$ Lawrence Berkeley National Laboratory, 1 Cyclotron Road, Berkeley, CA 94720, USA\\
}
}

       \date{Accepted 2023 November 02. Received 2023 October 24; in original form 2023 August 02}
       \pubyear{2023}

\begin{document}
%\linenumbers

\label{firstpage}
\pagerange{\pageref{firstpage}--\pageref{lastpage}} 
\maketitle

\begin{abstract}
  While proper orbital elements are currently available 
  for more than 1 million asteroids, taxonomical information
  is still lagging behind.  Surveys like SDSS-MOC4
  provided preliminary information for more than 100,000 objects,
  but many asteroids still lack even a basic taxonomy.
  In this study, we use Dark Energy Survey (DES) data to provide new
  information on asteroid physical properties. By cross-correlating
  the new DES database with other databases, we investigate
  how asteroid taxonomy is reflected in DES data. 
  While the resolution of DES data is not sufficient to distinguish between different asteroid
  taxonomies within the complexes, except for V-type objects,
  it can provide information on whether an asteroid
  belongs to the C- or S-complex.  Here, machine learning
  methods optimized through the use of genetic algorithms were
  used to predict the labels of more than 68\,000 asteroids with no prior
  taxonomic information.  Using a high-quality, limited set of asteroids
  with data on $gri$ slopes and $i-z$ colors, we detected 409 new possible
  V-type asteroids.  Their orbital distribution is highly consistent with
  that of other known V-type objects.

  \end{abstract}

\begin{keywords}
Minor planets, asteroids: general, catalogues, celestial mechanics.
\end{keywords}

\section{Introduction}
\label{sec: intro}

Currently, we know more than 1 million asteroids for which synthetic
proper elements can be reliably obtained using the method of
\citet{Knezevic_2003}.  Unlike osculating elements, proper elements
are constants of motion on timescales of Myr, which allows for the identification
of asteroid families. Our knowledge of the
physical properties of asteroids is, however, much more limited.  A full
spectral classification of asteroids
is available in various surveys for slightly more than 2000 objects
\citep{2002Icar..158..146B,2004Icar..172..179L, 2009Icar..202..160D}.  Preliminary taxonomical information can be obtained from surveys like the SDSS-MOC4
\citep{2001AJ....122.2749I}, using the method described in
\citet{2013Icar..226..723D}, for more than 100,000 asteroids.   More recent works on the matter are those of \citet{2018A&A...617A..12P} on the taxonomic classification of asteroids based on MOVIS near-infrared colors, and the new 3D machine learning classification scheme based on SDSS-MOC4 data of \citet{2022A&A...664A..51R}.
Yet, many objects, especially those at higher magnitudes and smaller diameters, lack any physical information.  This limits studies on asteroid
families, which are assumed to be mostly uniform in physical properties.

Asteroids can be classified into three main taxonomical groups based on their reflectance spectra.  Asteroids belonging to the C-complex are typically
dark in color and have low albedos (reflectivity), while S-complex
asteroids are typically brighter and more reflective. In the
\citet{2013Icar..226..723D} taxonomy, the X class is divided into three classes, E, M, and P, which are distinguished solely by their albedo (P $< $0.075, 0.075 $<$ M $<$ 0.30, E $>$ 0.30). V-type asteroids
are characterized  by a deep absorption band around 1 micrometer, which is
thought to be caused by the presence of the mineral olivine.  They are
thought to originate from the mantle of differentiated parent bodies.

The Dark Energy Survey (DES; \citealt{2005IJMPA..20.3121F}, see also \citet{2016MNRAS.460.1270D}) is used here to provide additional information on a set of more than 60,000 asteroids. DES is a collaborative effort that covered 5000 square degrees of the sky in the grizY bands from 2013 to 2019, primarily in the southern celestial sphere, aiming at investigating the dark energy. It is clear, however, its importance as also a Solar System survey. In fact, among its contributions, DES discovered and characterized a large and distant scattered disk object \citep{2017ApJ...839L..15G}, improved predictions of stellar occultations by numerous TNOs and Centaurs \citep{2019AJ....157..120B}, made hundreds of discoveries in the TNO region \citep{2022ApJS..258...41B}, provided a detailed photometric analysis of a large sample of Jupiter Trojans \citep{2022PSJ.....3..269P}, produced the largest TNO color and light curve catalog, facilitated the development of techniques to obtain optimal measurements of fluxes, colors, binarity, and variability for these slow-moving objects \citep{2023arXiv230403017B}, and even discovered a messenger from the outskirts of the Solar System \citep{2021ApJ...921L..37B}.

We used two sets of data from DES photometric measurements, a high-quality set
with $gri$ slopes and $i-z$ colors, where the classification method of
\citet{2013Icar..226..723D} can be applied, and a much larger
$(g-r,g-i)$ database.  We then cross-referenced the DES data with taxonomical, SDSS-MOC4, 
and albedo information, to understand how asteroid taxonomies are mapped
in the new dataset.  Based on the distribution of known asteroid
taxonomies, predictions on unlabeled bodies
can then be made, using machine learning methods, optimized
by the use of genetic algorithms \citep{chen_2004}.  Special attention is
then given to more robust taxonomical classification for
asteroid taxonomies that showed a good performance in DES data,
such as the important V-type asteroids.

In section 2, we describe how the data was obtained from the DES database and
organized. In section 3, we describe the use of the DeMeo and Carry taxonomy.
Studies using the $(g-r,g-i)$ data are done in section 4.
Section 5 presents DES candidates for V-type asteroids, and conclusions are
given in section 6. 

\section{Obtaining DES data}
\label{sec: des}

All colors presented in this study were obtained through observations and
measurements conducted by DES.

To search for known Solar System objects, we queried the entire 
DES database \citep{2021ApJS..255...20A} using keywords from image headers like
pointing coordinates, date and time of observations \citep{diehl2023dark},
exposure time, and filter \citep{flaugher2015dark}. We used an SQL-based
tool called {\it easyaccess} \citep{2018ascl.soft12008C} for all queries of
the DES database.

Having those pieces of information in hand, we
identify the single-epoch CCDs that could have captured the image of a
small Solar System body using the Sky Body Tracker
(SkyBoT; \citet{2006ASPC..351..367B}). SkyBoT, among other functionalities,
yields a list of all known Solar System objects within a given field of view
(FOV) when pointing coordinates, UTC date and time of observation, observing
site coordinates, and FOV angular size are provided. 
We selected objects whose dynamical classification, as provided by the SkyBoT,
were Hungaria, MB>Inner, MB>Middle, MB>Outer, MB>Cybele, MB>Hilda,
or Jupiter Trojan, where MB stands for Main Belt.

We obtained positions, magnitudes and other pieces of information of the selected objects from the Year-6 (Y6) list of objects so-called
\textsc{Y6A1\_FINALCUT\_OBJECTS}. The respective zero-points of each CCD were added to the magnitudes using the Forward Global Calibration Method \citep{2018AJ....155...41B}, prepared by the collaboration \citep[refer to Sects.~1 and 4.8 in][for more details about the finalcut catalog]{2018PASP..130g4501M}. 

Methods for improved photometry, applied to outer Solar System objects present in the images of DES, have been recently developed  \citep{2023arXiv230403017B}. However, the Y6 catalog has features that make it an extremely attractive and valuable source to a variety of photometric studies, the one presented here in particular: (i) the same methods/procedures are used to derive flux measurements over the whole survey area, allowing us to coherently correlate colors of objects from different sky regions and (ii) readily available (upon password) data, thus saving a lot of CPU time, in addition to (iii) high quality single epoch photometry. 

Table~\ref{Table: Raw numbers} shows the total number of asteroids that
SkyBoT identified as belonging to a DES frame. We selected objects classified
as Hungaria ($1.0 < a < 2.0$ au and $a(1-e) > 1.666$ au), Inner Main Belt
($2.0\leq a< 2.5$ au), Middle Main Belt ($2.5\leq a< 2.82$ au), Outer Main Belt
($2.82\leq a< 3.27$ au), (Main Belt) Cybele ($3.27\leq a< 3.7$ au), Hilda
($3.7\leq a < 4.6$ au) and Trojan ($4.6\leq a< 5.5$ au) according to the
SkyBoTclassification\footnote{\url{https://vo.imcce.fr/webservices/skybot/?documentation}}. However, not all of them were detected, often due to a faint
magnitude that is difficult to observe in a single-epoch image or to a very
large (many degrees in some cases) positional uncertainty. A brief description
of the task and tools used to find known Solar System objects in DES images
can be found in the work by \citet{2019AJ....157..120B}. All positions and
magnitudes were obtained from the Y6 final cut DES catalogs and were
queried using the \textit{easyaccess} tool. 

It is important to note that the DES observational cadence may not always be
suitable for determining the colors of small objects, since observations of a
same object in different filters may be separated by long periods of time and
we do not have accurate enough rotational information to correct for rotational
effects. As a result, we formed colors for a given object using only
observations that were obtained within 10 minutes of each other
\citep{abbott2021dark}.

If we were to conduct an observational run specifically dedicated to
measuring the colors of small objects, 10 minutes would be a reasonable
estimate for obtaining magnitudes in different filters, taking into account
the exposure time, readout and filter change. However, a very fast rotator
would likely require simultaneous observations in the different filters.

The ALCDEF \citep[Asteroid Light Curve Data Exchange Format, see][]{2018DPS....5041703S,2011MPBu...38..172W,2010DPS....42.3914S} website
displays a plot\footnote{\url{https://alcdef.org/php/alcdef_aboutLightcurves.html}} indicating that most asteroids have rotational periods longer than 2 hours,
with many falling between 4 and 10 hours. If we consider the shortest period of 2 hours, 10 minutes of observations corresponds to a rotation of 30 degrees, while for a 7-hour period, we have a rotation of only 8.6 degrees. Therefore, our decision to group observations of the same object that were acquired within 10 minutes appears to be a good compromise between minimizing the impact of rotational effects on colors while keeping a sample as large as possible.

We can estimate the maximum error caused by neglecting pure rotational
effects (i.e., considering a surface with homogeneous albedo)
in our database. If we simplify the asteroid light curves as a
double-peaked triangle wave and assume a large amplitude,
of 0.5 mag for a 4-hour rotation period, a 10-minute separation
between observations corresponds to a magnitude change of aproximately
$10/(4 \cdot 60/4) \cdot 0.5 \simeq 0.083$. For a
10-hour rotation period, the magnitude change would be only 0.033.
Fast-rotating asteroids ($\simeq 4$ hr period) would generally have smaller
amplitudes because elongated  objects are more prone to breaking (see Figure 9
in \citet{Chang_2019}).  In most cases, the rotational effect on 
color should be of the order of $10^{-2}$, and should not exceed 0.1.

\begin{table}
  \begin{center}
    \caption{Known Main Belt, Hungaria and Trojan objects in DES images as indicated by the SkyBoT. Note that the number of objects effectively present in the images is smaller.}
    \label{Table: Raw numbers}
    \begin{tabular}{|l|c|}\toprule
\hline
Class & Number of objects \\
\hline
Hungaria & 12\,199 \\
Main Belt - Inner  & 102\,979 \\
Main Belt - Middle & 134\,277 \\
Main Belt - Outer  & 125\,823 \\
Cybele & 1\,876 \\
Hilda  & 1\,751 \\
Trojan & 1\,684 \\
\hline
\end{tabular}
\end{center}
\end{table}

Albedo variations, however, do exist and have statistical significance
on color variations (see discussion in \citet{2004MNRAS.348..987S}).
Its overall impact can be inferred from table~(\ref{Table: Szabo}).  The
color of a given object (e.g., $g-r$) is determined by the mean value of
multiple measurements of that color and its uncertainty is represented by
the respective standard deviation.  One might reasonably anticipate that
color uncertainties in color depend on the number of measurements under
the influence of a variable albedo (the smaller the average over a rotation,
the more probable it is to obtain a large uncertainty).
Table~(\ref{Table: Szabo}) shows that this is not the case.  Based on this
analysis, it can be concluded that albedo variations on the surface of
asteroids are not likely to introduce significant systematic effects on our
results.
  
\begin{table}
  \begin{center}
    \caption{DES $(g-r)$ colors for inner main belt asteroids.
      We report the standard deviations of measurements in $(g-r)$, the number
      of colors recorded per object, and the total count of $(g-r)$
      measurements used to calculate the mean value. Additionally.
      we present the total count of objects with a $(g-r)$ color, regardless
      whether they were utilized in this study.}
    \label{Table: Szabo}
    \begin{tabular}{|c|c|c|}\toprule
\hline
Stand. Dev in & \# of $(g-r)$       & Total number of $(g-r)$      \\
$(g-r)$        & colors per object & colors used                 \\
\hline
   0.041     &         1 to 3       &           20825 \\            
   0.058     &         4 to 6       &           403   \\            
   0.059     &         7 to 9       &           30    \\            
   0.064     &        10 to 12      &           32    \\           
   0.051     &        13 to 15      &           7     \\            
   0.028     &        16 to 18      &           10    \\            
   0.044     &        19 to 21      &           5     \\            
   0.042     &        22 or more    &           7     \\            
\hline
\end{tabular}
\end{center}
\end{table}

Finally, in order to compare our data with known sources in the literature on
asteroid taxonomy (see Sects. 3 and 4), we transformed DES colors into
SDSS (Sloan Digital Sky Survey) ones using the transformation equations
from \citet{2021ApJS..255...20A}.

However, instead of using individual equations for each filter, we used
their differences. In this context, to provide the necessary photometric
information for section 3 and 4, we only needed the following colors from
DES: $(g-r)_{\rm DES}$, $(g-i)_{\rm DES}$, $(r-i)_{\rm DES}$ and $(i-z)_{\rm DES}$. Solar apparent magnitudes and central wavelengths in the SDSS, where necessary, were taken from \citet{2018ApJS..236...47W}. More specifically, the DES to SDSS color transformations used were

\begin{equation}\label{eq:des2sdss}
\begin{split}
    (g-r)_{\rm SDSS}&=(g-r)_{\rm DES}+0.060\times(g-i)_{\rm DES}\\&-0.150\times(r-i)-0.019 \\
    (g-i)_{\rm SDSS}&=1.060\times(g-i)_{\rm DES}-0.167\times(r-i)_{\rm DES}\\&+0.022 \\
    (i-z)_{\rm SDSS}&=(i-z)_{\rm DES}+0.113\times(r-i)-0.003
\end{split}
\end{equation}
, where the first members of Eqs.~\ref{eq:des2sdss} are
SDSS colors and colors in the second members are from DES.
The variances ($\sigma^2$) in SDSS colors are given by the error propagation of these equations, added quadratically the root mean square of the respective band tranformations from \citet{2021ApJS..255...20A} used to determined each of the color equations above.

\section{DES database: DeMeo and Carry taxonomy}
\label{sec: demeo_tax}

As a preliminary step of our analysis, we selected asteroids from the
DES database for which both $gri$ slopes and $i-z$ colors
are both available. The color indices contain information that can
be used to derive a very low-resolution reflectance spectrum denoted
as $Rf$.  We calculate $Rf$ using the equation $Rf= 10^{-0.4(M_{f}-M_{\odot f})}$,
where $M_{f}$ is the magnitude of the asteroid and $M_{\odot f}$
is the magnitude of the Sun in the same filter. To normalize $Rf$ to 1 at a
given wavelength we use the relationship $Rf,v = 10^{-0.4[(M_f-M_v)-(M_{\odot f}-M_{\odot v})]}$.
The spectral gradient, or slope, is then given by (with $\lambda$ in nanometers, see also \citet{2009A&A...494..693S}):

\begin{equation}
  S({\lambda}_1,{\lambda}_2)=10^{4}\frac{Rf_{2,v} - Rf_{1,v}}{|{\lambda}_2-{\lambda}_1|}.
  \label{eq: spect_gradient}
\end{equation}
In our case, $v$ is the SDSS $g$ filter.

This slope, as given above, is expressed as a percentage per 100 nanometers. To calculated the $Rf,v$ value for each asteroid, we used the Sloan  $g$, $r$, and $i$ magnitudes and the effective wavelengths \citep[see][]{2018ApJS..236...47W} of each of
these filters and then fit a straight line to the pairs $(\lambda,Rf,v)$. The angular coefficient from this adjustment gives the $gri$ slope in \% / 100nm.

\begin{figure}
  \centering
  \centering \includegraphics[width=3.0in]{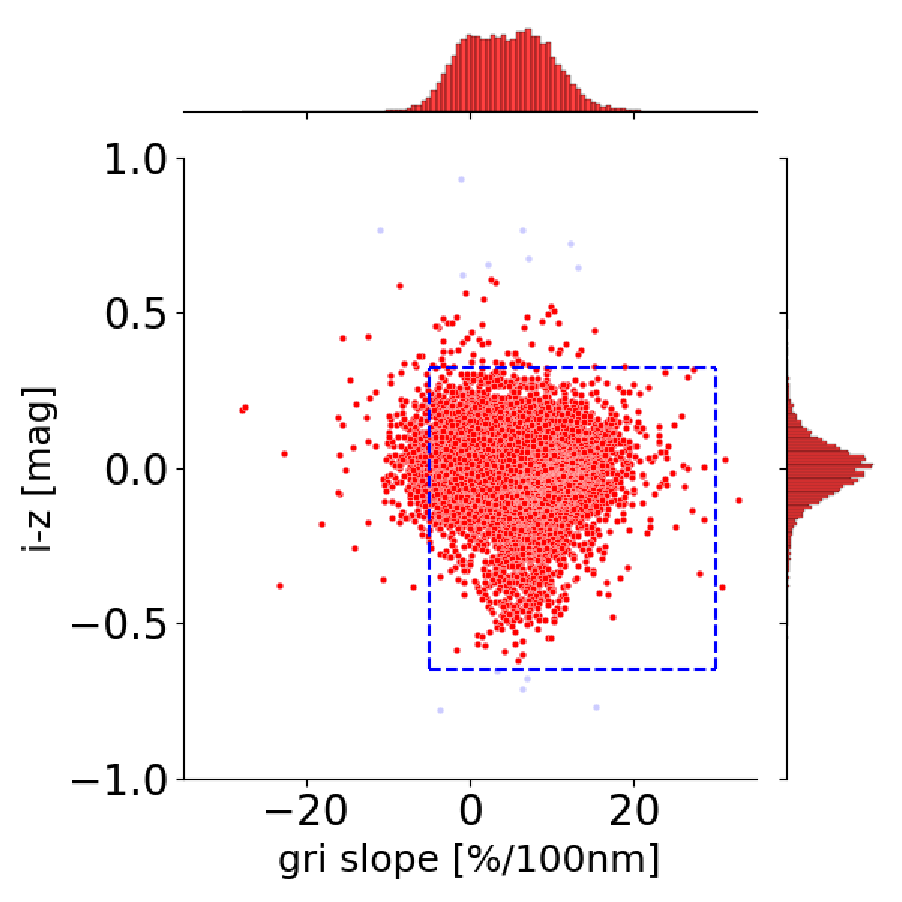}

  \caption{A joint plot of asteroids with
    available DES data, showing their $gri$ slope,$i-z$) values
    (blue transparent full circles). The full red
    circles represent asteroids with values of the two parameters within
    the 0.15 to 0.85 quantile interval in 
    the $gri$ slope and $i-z$ color distributions. The
    dashed blue box displays the interval for which the
    \citet{2013Icar..226..723D} classification scheme applies. The top and right
    side of the figure show histograms of the $gri$
    slope and $i-z$ colors distributions without outliers, respectively.}
\label{Fig: DES_prel}
\end{figure}

These parameters are necessary to classify the asteroid
taxonomies according to the method of \citet{2013Icar..226..723D}, which
is based on data from the Sloan Digital Sky Survey-Moving Object Catalog data
(SDSS-MOC4; \citet{2001AJ....122.2749I}).  Our objective here is to identify
asteroids present in both catalogs and to infer what taxonomical properties
can be obtained from DES data.  Given their temporal constraints, 17154 asteroids with $gri$ slopes and $i-z$ colors are present in the DES database.
To eliminate outliers, we only consider asteroids within the 0.15 to 0.85
quantile interval for both parameter distributions\footnote{Quantiles
are cut points that divide the range of a probability distribution into
continuous intervals with equal probabilities.  The median value of a
distribution would correspond to a quantile of 0.50.  More information
on the procedure to compute quantiles can be found in
\citet{blitzstein2019introduction}.}.  This corresponds to intervals in
$gri$ slope between -34.738 and 43.466, and in $i-z$ between -0.627 and 0.612.
Outliers are displayed as blue full transparent circles
in figure~(\ref{Fig: DES_prel}).  For
statistical reasons, we will exclude outliers from our analysis, hereafter,
yielding a sample of 17135 asteroids, 10685 of which are numbered, and the rest
multi-opposition asteroids.  Most of the outliers are data points beyond
the range for which the \citet{2013Icar..226..723D} method applies, shown
as a dashed blue box in figure~(\ref{Fig: DES_prel}),
and most of them are not shown in figure~(\ref{Fig: DES_prel}).
Therefore, removing these objects does not cause any significant loss of
information\footnote{Other quantile intervals were also considered.  For a
distribution within the 0.10 to 0.90 quantile interval only 2 outliers
were found. This number increases to 42 if we consider distributions within
0.20 and 0.80. However, for such distributions, some data points within the
range for which the classificatiom method proposed by
\citet{2013Icar..226..723D} applies where also
excluded. For these reasons, we decided to work with the 0.15 to 0.85
quantile interval.}. The range of absolute magnitudes for the data set
without outliers goes from 9.0 to 20.5.  On the top and right side of
figure~(\ref{Fig: DES_prel}), we display
histograms of $gri$ slope and $i-z$ colors for the population of
asteroids without outliers.  While there is a single peak for the $i-z$
distribution, neither distribution is normal.  If we compute the skewness,
which is a measure of the symmetry or asymmetry of a distribution, and the
kurtosis, which measures whether the data are heavy-tailed or light-tailed
relative to a normal distribution, both distribution moments are
significantly different from 0, which is the expected value for a Gaussian
distribution. Both are skewed, one toward the right and
one toward the left flank, with values of skewness of 0.14 and -0.53,
respectively. Both are leptokurtic distributions, with heavier tails than a
normal distribution. The values of kurtosis are 3.31 and 6.45, respectively.
These results show that neither of the distributions can be modelled as
Gaussian distributions, and that more complex models should be
used for this data.

\begin{figure*}
  \centering
    \begin{minipage}[c]{0.45\textwidth}
    \centering \includegraphics[width=2.2in]{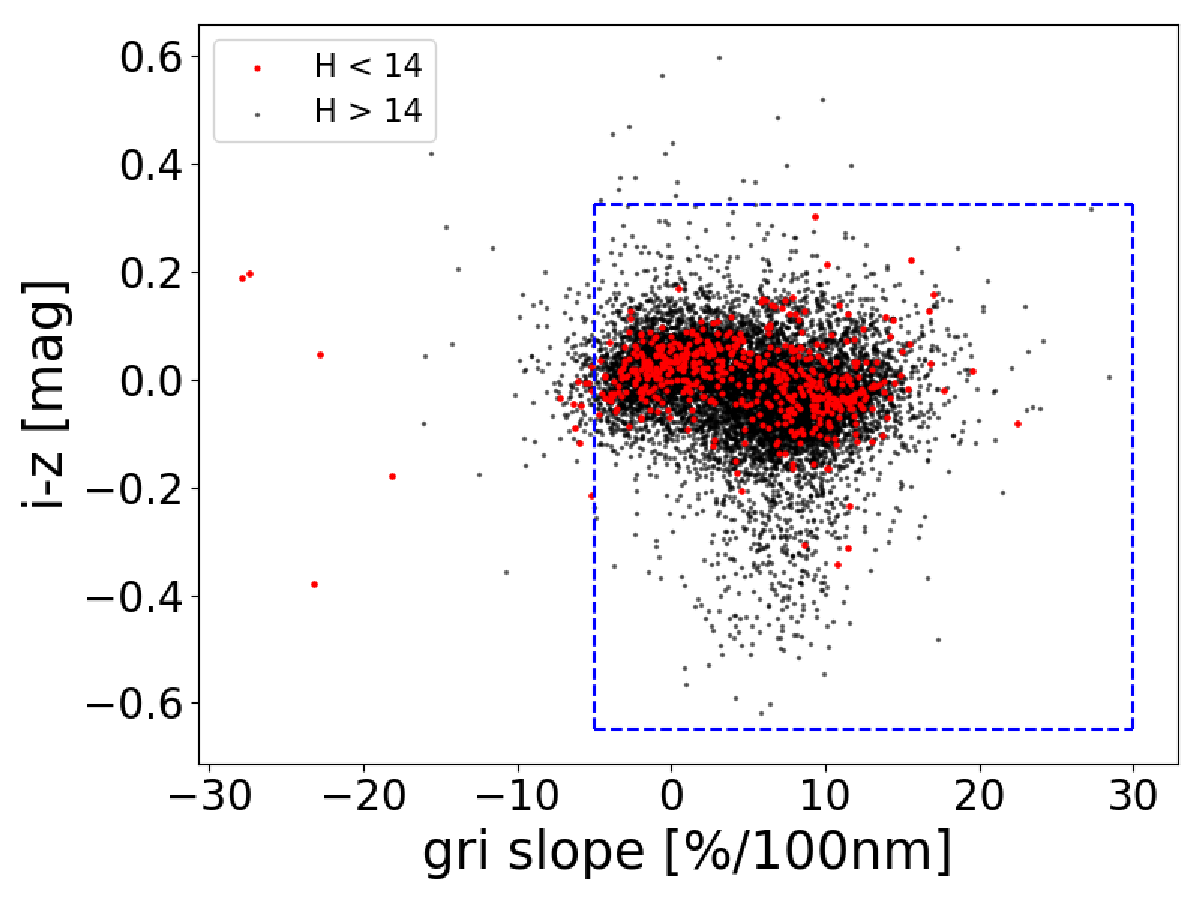}
  \end{minipage}%
  \begin{minipage}[c]{0.45\textwidth}
    \centering \includegraphics[width=2.2in]{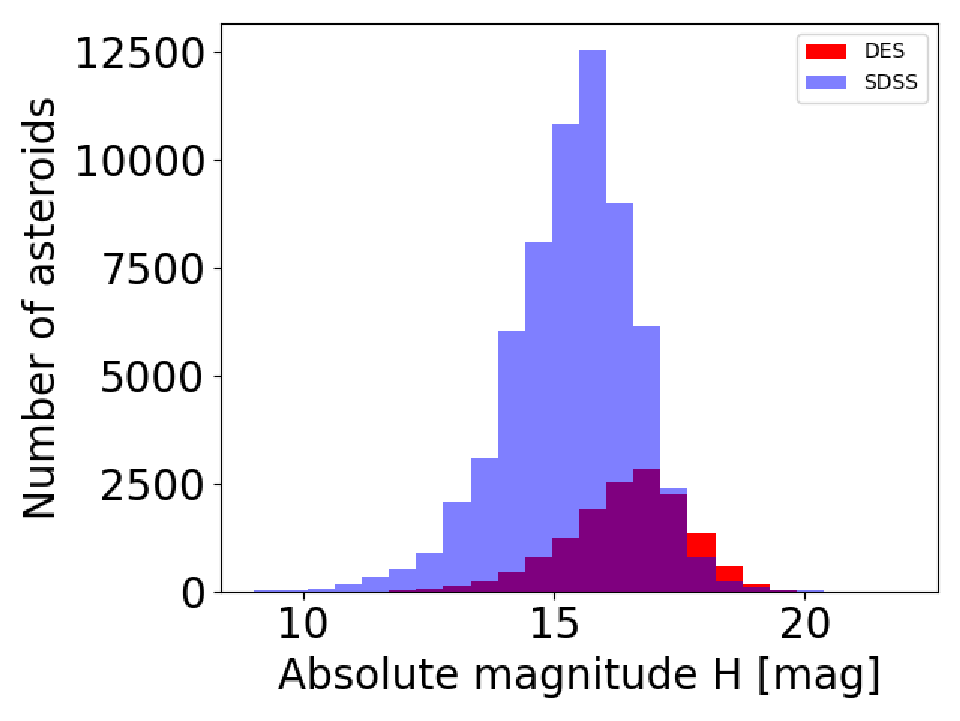}
  \end{minipage}
  \caption{Left panel: a projection in the ($gri$ slope,$i-z$) plane
    of DES large ($H < 14$) and small ($H > 14$) asteroids. The distribution
    of the two populations is not significantly different.  In the right panel, we show the absolute magnitude distribution of asteroids
    in the SDSS-MOC4 database and in the DES data set, for which $gri$ slope
    and $i-z$ colors are available.  The DES survey covers fainter objects
    than those reached by the SDSS one.}
\label{fig: DES_H_Check}
\end{figure*}

Having obtained the data set of asteroids in the ($gri$ slope,$i-z$) plane,
it may be important to check how properties like the asteroids' absolute
magnitudes may correlate with these data. While the size of an asteroid
also depends on its albedo, objects with small absolute magnitudes
tend to have larger sizes, and vice versa. In the main belt, more than
40\% of the asteroid population belong to asteroid families
\citep{2014Icar..239...46M}, and the distribution of large and small objects
tends to be not significantly different in proper element domains
\citep{2017A&A...598A..52G}.
This is also observed in the
($gri$ slope,$i-z$) plane, as shown in the left panel of
figure~(\ref{fig: DES_H_Check}),  where the distributions of large and small objects are fairly similar.  While smaller objects cover a larger area than thelarger ones, this is likely due to their greater number. Regions with a high number density of asteroids are inhabited by both small and large objects.  Also, DES data observed fainter
objects than previous surveys, like the Sloan Digital Sky Survey-Moving Object
Catalog data (SDSS-MOC4 or SDSS, for brevity; \citet{2001AJ....122.2749I}).
The right panel of figure~(\ref{fig: DES_H_Check}) shows histograms of
absolute magnitude distributions for the two surveys.   Absolute
magnitude $H$ were obtained from the Asteroid Families Portal $AFP$
("http://asteroids.matf.bg.ac.rs/fam/index.php", \cite{Radovic_2017},
accessed on June 2023). Nominal errors on $H$ are of the order of
one decimal digit \citep{2012Icar..221..365P} \footnote{The same authors found that there is a systematic negative offset of absolute magnitudes in
catalogs, which reaches a peak of -0.5 around $H = 14$.  We are not correcting
for these biases in this work, since our focus is on larger values
of $H$.  However, it is important to alert the reader to
these biases for data around $H \simeq 14$.}.
We expect that the new data on fainter objects observed
by DES could provide new insights on the physical properties of small
asteroids.

\begin{figure}
  \centering
  \centering \includegraphics[width=3.0in]{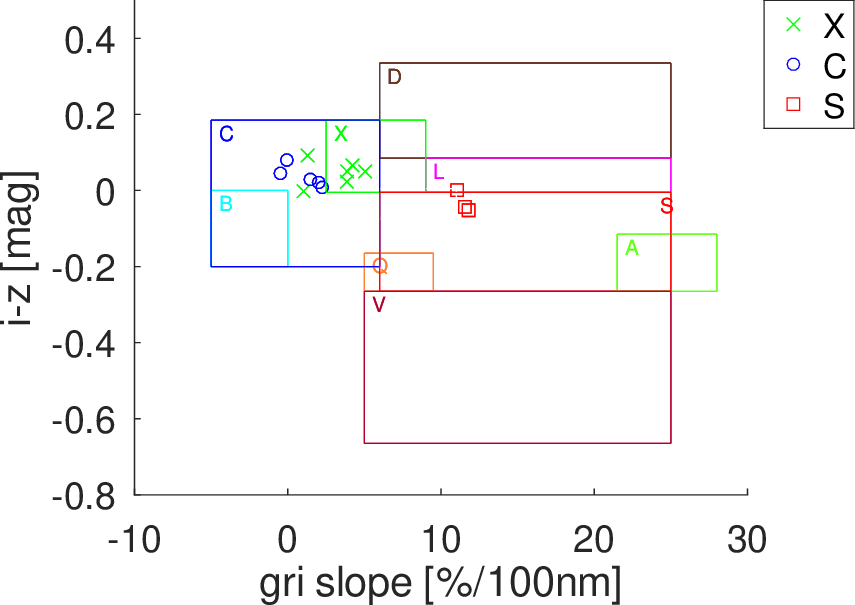}
  \caption{$i-z$ colors versus $gri$ slope for asteroids
    with taxonomical data present in the DES database. The colored boxes
    represent the boundaries of the \citet{2013Icar..226..723D} taxonomical
    classes.}
\label{Fig: Tax_1}
\end{figure}

To begin, we aim at finding location of asteroids with known
taxonomies in this new data set. For this purpose, we refer to the 
asteroid taxonomical data available in the surveys
of \citet{2002Icar..158..146B}, \citet{2004Icar..172..179L}, and
\citet{2009Icar..202..160D}.  For asteroids with more than one entry
in the three surveys, we use a majority
vote method to assign the most likely spectral type.
There were 14 asteroids with taxonomical data in our
selected DES data: 5 C-type, 3 S-type, and
6 X-type.  Their location in the
$(gri, i-z)$ diagram is shown in figure~(\ref{Fig: Tax_1}).  Apart from two
X-type asteroid, there is no indication that the DES parameters are inconsistent with the classification scheme of \citet{2013Icar..226..723D}.
However, small number statistics
prevent us from reaching more compelling conclusions.  Most of the errors
that we observed are in the horizontal axis, which are associated to errors
in the $gri$ slopes.

\begin{figure*}
  \centering
    \begin{minipage}[c]{0.45\textwidth}
    \centering \includegraphics[width=3.1in]{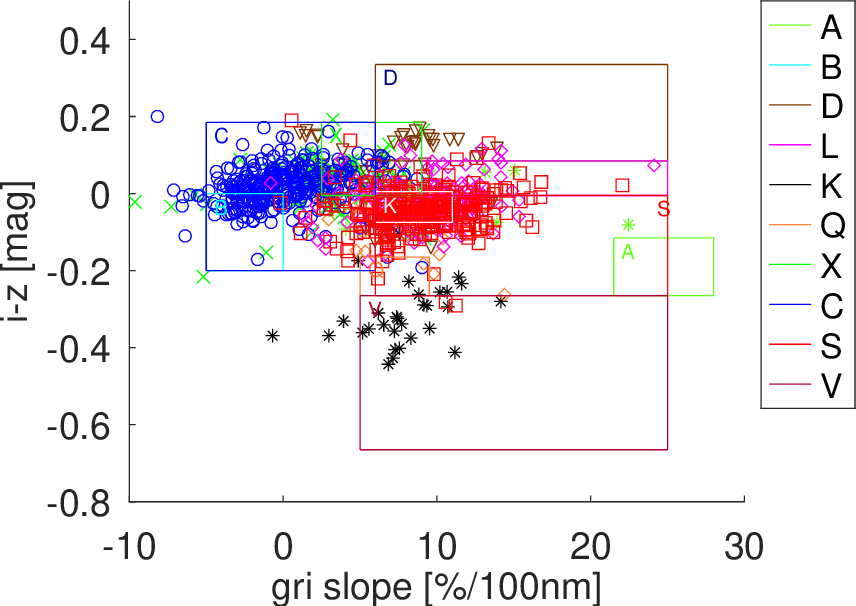}
  \end{minipage}%
  \begin{minipage}[c]{0.45\textwidth}
    \centering \includegraphics[width=3.1in]{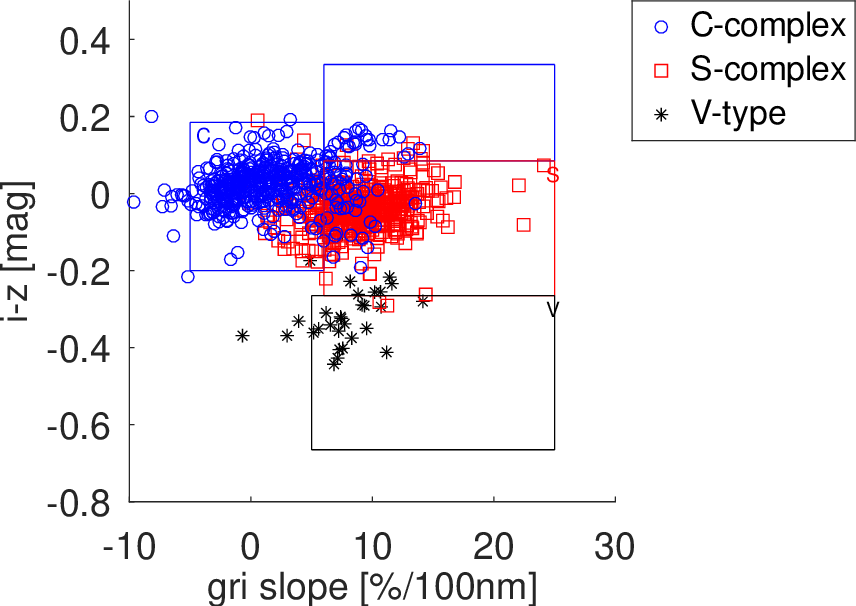}
  \end{minipage}
  \caption{$i-z$ color versus $gri$ slope of asteroids
    with data in both the SDSS-MOC4 and DES database.  The boundaries
    of the \citet{2013Icar..226..723D} taxonomical classes are identified
    by the colored boxes.  The left panel displays the positions of these
    asteroids for all \citet{2013Icar..226..723D} available classes,
    while the right panel shows a simplified classification
    scheme that only identifies C-complex, S-complex asteroids, and V-type
    objects.}
\label{fig: DES_SDSS}
\end{figure*}

To increase the number of asteroids with taxonomical information, we turned
our attention to the Sloan Digital Sky Survey-Moving Object Catalog data
(SDSS-MOC4; \citet{2001AJ....122.2749I}). Although the taxonomical information
from the SDSS-MOC4 data has some limitations, as it is based on photometrical
colors, it can still provide useful preliminary information on
asteroids' physical properties.  Firstly, we eliminate asteroids with
large errors in $gri$ slope (error larger than 10 \%/100nm) and
$i-z$ color (error larger than 0.1 mag). Then, we identify 950 asteroids for
which the method of \citet{2013Icar..226..723D} can be used to obtain asteroid
taxonomies.  Results for the whole SDSS-MOC4 data are available at
https://sbn.psi.edu/pds/resource/sdsstax.html. The identified asteroids
include: 100 X-types, 54 D-types, 345 C-types, 151 L-types, 29 Q-types,
271 S-types, 5 A-types, and 32 V-types. We neglected subclasses like
the CX, SQ, SV, LS and QV, since we will show that DES data does not have the
resolution needed to perceive these subtle differences.  SV and QV objects
were classified as V-type since they are all found in regions of
the ($gri$ slope, $i-z$) plane occupied by this class of objects.
The left panel of figure~(\ref{fig: DES_SDSS}) displays the position in
the plane of $i-z$ color versus $gri$ slope for all these asteroids.

The classification obtained from DES data does not
always agree with that from SDSS-MOC4, also because the usually large
uncertainties on DES $gri$ slopes.  To further check the validity of the
DES taxonomy, we alo performed correlations with the
\citet{2010A&A...510A..43C} and \citep{2018A&A...617A..12P} data sets.
Table~(\ref{Table: Classification_accuracy}) presents the classification
accuracy for asteroids in various spectral types for cross-correlations
with the three data sets: C-types and V-types
can be classified by DES data with purities, defined as
the fraction of DES taxonomic labels correctly classified as such,
as of 50\% or higher, while
the other types have low purity percentages.
Interestingly, 44.4\% of
the misclassified V-type are found in the central and outer main belt,
while no confirmed V-type has been found in these regions.  This suggests that
V-type candidates identified from DES data in these areas should be approached
with caution.  SV types can be
easily misclassified as V-types using DES data, and none of the A-types
were classified correctly.  Based on this analysis, we propose
a limited DES taxonomical classification consisting of three groups,
the C- and S-complex, and the V-types.
The C-complex will include X, D, and C-types.
while the S-complex will encompass the remaining A, L, Q, K, and S-types.
The V-types will remain as a separate class.

\begin{figure}
  \centering
  \centering \includegraphics[width=3.0in]{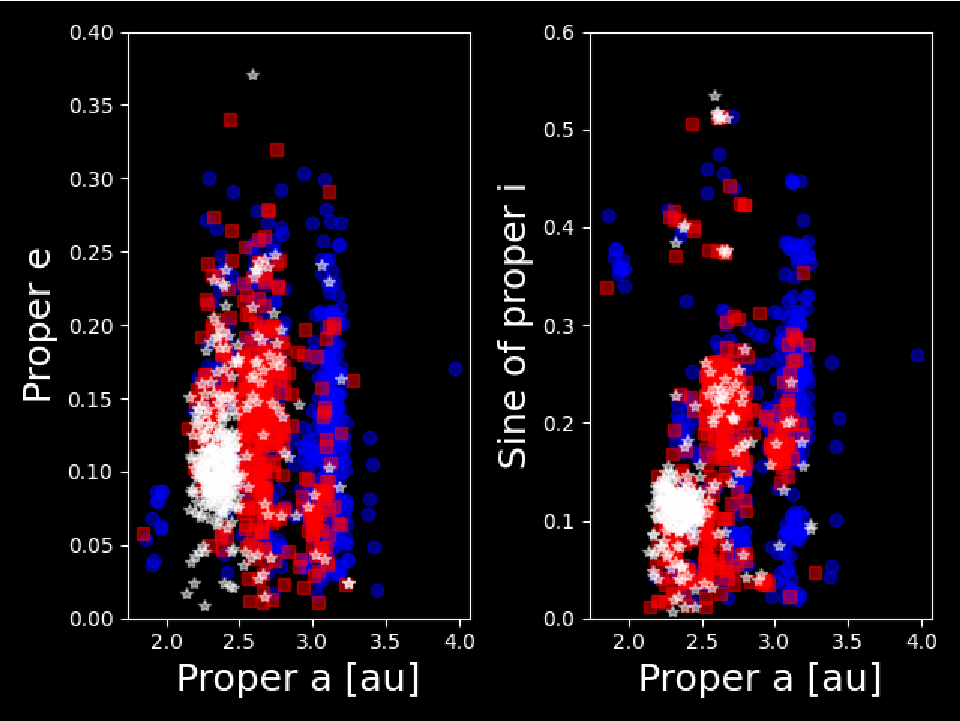}
  \caption{Projections in the proper $(a,e)$ and proper $(a,\sin{(i)})$ of
    the C-complex (blue circles), S-complex (red squares) and V-type objects
    (white asterisks), identified in both the DES and SDSS-MOC4 databases,
    and as listed in table~(\ref{Table: Classification_accuracy_rev}).
    Proper elements were obtained from $AFP$.}
  \label{Fig: Ivezic_DES}
\end{figure}

The new results for DES data are presented in the right panel of 
figure~(\ref{fig: DES_SDSS}) and
table~(\ref{Table: Classification_accuracy_rev}).  The purities
for both complexes are now above 75.0\%, suggesting that
this simplified scheme could be applied more successfully.
Following the analysis of \citet{2002AJ....124.2943I}, we also plot in
figure~(\ref{Fig: Ivezic_DES}) the orbital distribution of C-complex,
S-complex, and V-type objects as identified in both DES and SDSS-MOC4.
Proper elements were obtained from $AFP$.
As expected, C-complex asteroids are more common in the outer main belt,
while S-complex asteroids are mostly found in the inner main belt, with some
mixing of the two complexes.  We can easily identify single asteroid families,
like the S-complex Eunomia family in the central main belt, roughly
at $a \simeq 2.6$ au, $\sin{(i)} \simeq 0.25$, or the Koronis family
in the pristine region, at $a \simeq 2.9$ au, $\sin{(i)} \simeq 0.05$.
V-type asteroids are mostly associated with the Vesta family at
$a \simeq 2.35$ au, $\sin{(i)} \simeq 0.12$, but are also found
in the central and outer main belt.  Because of the importance
of V-type asteroids for early scenarios of the Solar System formation, we
will further discuss their orbital distribution in
section~(\ref{sec: DES_V_type}).

\begin{table*}
  \begin{center}
    \caption{Percentage of consistent taxonomic classifications, or purity,
        for asteroids in both the DES and SDSS-MOC4, the DES and the
        \citet{2010A&A...510A..43C}, and the DES and the MOVIS databases.}
    \label{Table: Classification_accuracy}
    \begin{tabular}{|c|c|c|c|c|c|c|}\toprule
\hline
Ast. & \# of     & Percentage of SDSS-MOC4 &    \# of     & Percentage of Carvano et al. (2010) &  \# of     & Percentage of MOVIS   \\
type & asteroids & cons. classification & asteroids & cons. classification  & asteroids & cons. classification\\
\hline
X  & 100 & 41.0 & 91 & 40.7 &  9 & 22.2 \\
D  &  54 & 37.0 & 40 & 30.0 &  8 & 12.5 \\
C  & 345 & 63.8 &335 & 64.2 & 21 & 66.7 \\
A  &   5 &  0.0 &  5 &  0.0 & 65 &  0.0 \\
L  & 151 & 19.2 &150 & 18.8 & 28 & 32.0 \\
Q  &  29 & 34.5 & 33 & 30.3 &  0 &  -   \\
S  & 268 & 38.8 &270 & 38.5 & 55 & 34.5 \\
V  &  32 & 65.6 & 25 & 64.0 & 21 & 52.4 \\
\hline
\end{tabular}
\end{center}
\end{table*}

\begin{table}
  \begin{center}
    \caption{Percentage of consistent taxonomic classifications (purities)
      for asteroids in both the DES and SDSS-MOC4 databases for a revised
      taxonomical scheme.}
    \label{Table: Classification_accuracy_rev}
    \begin{tabular}{|c|c|c|}\toprule
\hline
Asteroid   & \# of     & Percentage of       \\
complex & asteroids & cons. classification \\
\hline
C  & 499 & 94.2 \\
S  & 454 & 80.0 \\
V  &  32 & 65.6 \\
\hline
\end{tabular}
\end{center}
\end{table}

\begin{figure}
  \centering
  \centering \includegraphics[width=3.0in]{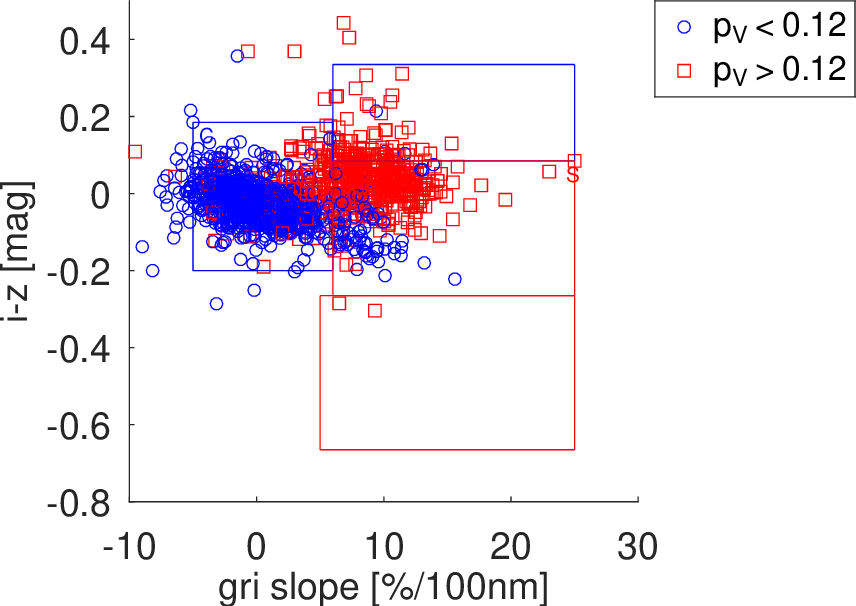}
  \caption{$i-z$ color versus $gri$ slope for asteroids
    with albedo data and present in the DES database. }
\label{Fig: WISE_DES_zi}
\end{figure}

Finally, we searched for objects in the DES dataset with albedo data in
the WISE and NEOWISE, AKARI, or IRAS databases
\citep{2012ApJ...759L...8M, 2011PASJ...63.1117U, 2010AJ....140..933R},
and we identified 1573 asteroids, as shown in figure~\ref{Fig: WISE_DES_zi}.
Albedo data correlates rather well with the position of asteroids in the
plane of $i-z$ color vs $gri$ slope:  81.5\% of objects identified
as C-complex have values of $p_V < 0.12$, and 94.6\% of S-complex
asteroids (we include V-type asteroids as S-complex bodies for albedo
purposes) have $p_V > 0.12$, as expected for these bodies.  The lower
value of C-complex asteroids with low albedo data, in contrast to
the higher percentage of S-complex asteroids with high albedos can be
attributed to the fact that C-complex asteroids include X-type, which, as
discussed in section~(\ref{sec: intro}), also include the M- and E-types,
which have high albedos. Table~(\ref{Table: Classification_pv}) summarizes our findings.

\begin{table}
  \begin{center}
    \caption{Percentage of asteroids with C and S-complexes classifications
    with geometric albedo $p_v < 0.12$ and $p_V > 0.12$, respectively.}
    \label{Table: Classification_pv}
    \begin{tabular}{|c|c|c|}\toprule
\hline
Asteroid & \# of     & Percentage of  \\
complex & asteroids & albedo values  \\
\hline
C  & 1188 & 81.5 \\
S  &  385 & 94.6 \\
\hline
\end{tabular}
\end{center}
\end{table}

\begin{figure}
  \centering
  \centering \includegraphics[width=3.0in]{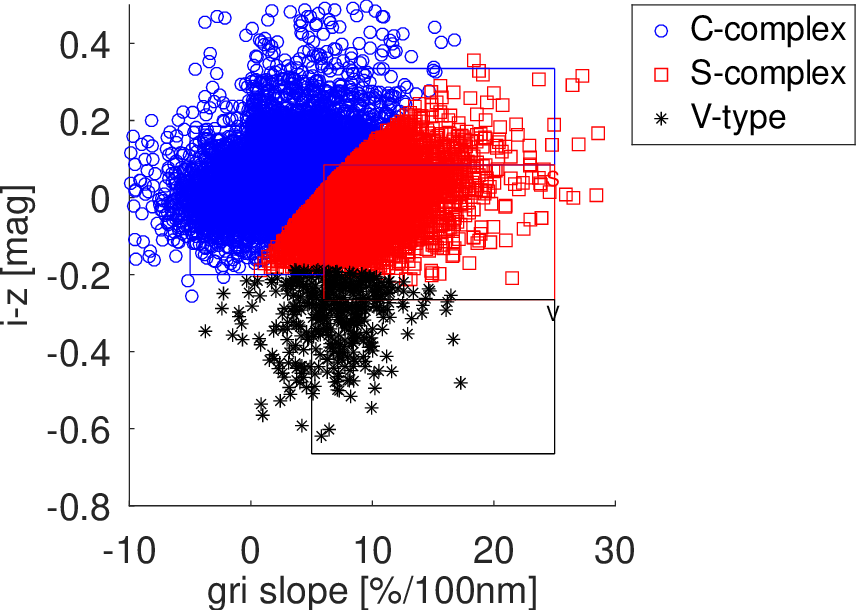}
  \caption{$i-z$ color versuss $gri$ slope of the ML predicted data for asteroids in the DES database with no SDSS-MOC4 data.}
\label{Fig: DES_GEN}
\end{figure}

What predictions can be made based on these results?  Based on the results
of the SDSS-MOC4 analysis, we can train machine learning (ML) algorithms.
In order to select the best-performing ML methods and the combination
of its free parameters, or hyper-parameters, that work best for our dataset,
we use genetic algorithms \citep{chen_2004}, according to the procedure
described in \citet{2021CeMDA.133...24C}.  We run the genetic algorithm
procedure on a subset of asteroids with both DES and SDSS-MOC4 data
three times, using a validation set corresponding to 20\% of the training
data.   The use of a validation set is recommended to avoid the issue
of overfitting, which happens when the model is overly sensitive
to the fine details of the training set, but may perform poorly when
dealing with other sets of data. The best-performing algorithm was a Gaussian
Naive Bayes (GNB) estimator \citep{10.5555/892239}. GNB is a classification
technique used in ML based on the probabilistic approach and Gaussian
distribution. GNB assumes that each parameter (also called features or
predictors) has an independent capacity of predicting the output variable.
The combination of the prediction for all parameters is the final prediction,
that returns a probability of the dependent variable to be classified in each
group. The final classification is assigned to the group with the higher
probability.  Finally, since we are dealing with a multi-classification
problem, the use of a $f_1$ score, defined as a harmonic
of precision and recall
($f_1 = 2 \cdot \frac{Precision \cdot Recall}{Precision + Recall}$),
is more appropriate to correctly estimate the efficiency of the algorithm,
rather than the more commonly used accuracy.  The model had an a $f_1$
score of 91.6\%.

Figure~(\ref{Fig: DES_GEN}) displays the results of a prediction for 16516
asteroids with no previous labels.  Predictions were made separately for
numbered and multi-opposition asteroids, using the same training data. The diagonal line in figure~(\ref{Fig: DES_GEN}) is caused by the presence of D-type objects that extends the boundaries of C-complex upward and rightward.  In the ML model, this causes the boundary between C-complex and S-complex to be a diagonal line.
Our classifier predicts that 10213
of the asteroids are likely to be C-complex, 5890 are likely S-complex
asteroids, and 410 are new possible V-type objects.

\section{DES database: (g-r,g-i) plane}
\label{sec: des_data}

\begin{figure}
  \centering
  \centering \includegraphics[width=3.in]{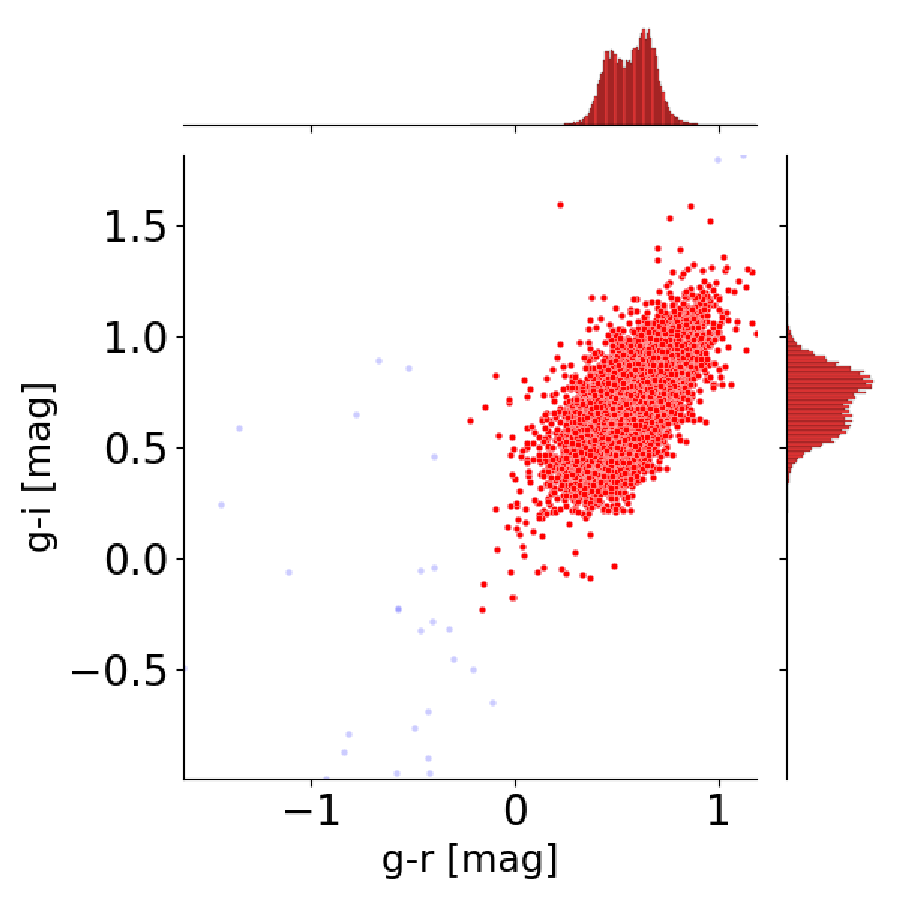}
  \caption{A joint plot of the $(g-r,g-i)$ colors for asteroids with available
    DES data. The blue full transparent circles indicate outlier data in the
    distributions of both colors.  The histograms of the two colors
    distributions, without the outliers, are displayed on the top and right
    sides, respectively.}
\label{Fig: DES_gen}
\end{figure}

The number of asteroids with DES data increases significantly if
we consider the $g-r$ and $g-i$ colors instead of the $i-z$ and
$gri$ slope, passing from 17078 to 61493.
Their distribution in the $(g-r,g-i)$ plane is shown in
figure~\ref{Fig: DES_gen}, with $g, r, i$ being the photometric bands
of the SDSS $'ugriz'$ system.  
While some of the brightest objects can have
negative values of these colors, most asteroids are found in an interval
from 0.2 to 1.2 in both quantities.  To eliminate these outliers, we again only
consider asteroids within an interval of quantiles of 0.15 to 0.85 for
both color distributions, which corresponds to the interval
$-0.266 < g-r < 1.393$ and $-0.326 < g-i < 1.746$.
Outliers are displayed as blue full circles
in the figure, while red full circles show the positions of the rest
of the dataset. We will exclude outliers from our analysis, henceforth.
After removing the outliers, we ended up with a dataset of 61142 asteroids. 
Figure~\ref{Fig: DES_gen} shows a joint plot and histograms of $g-r$
and $g-i$.  Both distributions
are single-peaked, slightly skewed, with skewness of -0.17 and -0.16,
respectively, and leptokurtic, with kurtosis values of 3.37 and 3.08.

\begin{figure}
  \centering
  \centering \includegraphics[width=3.0in]{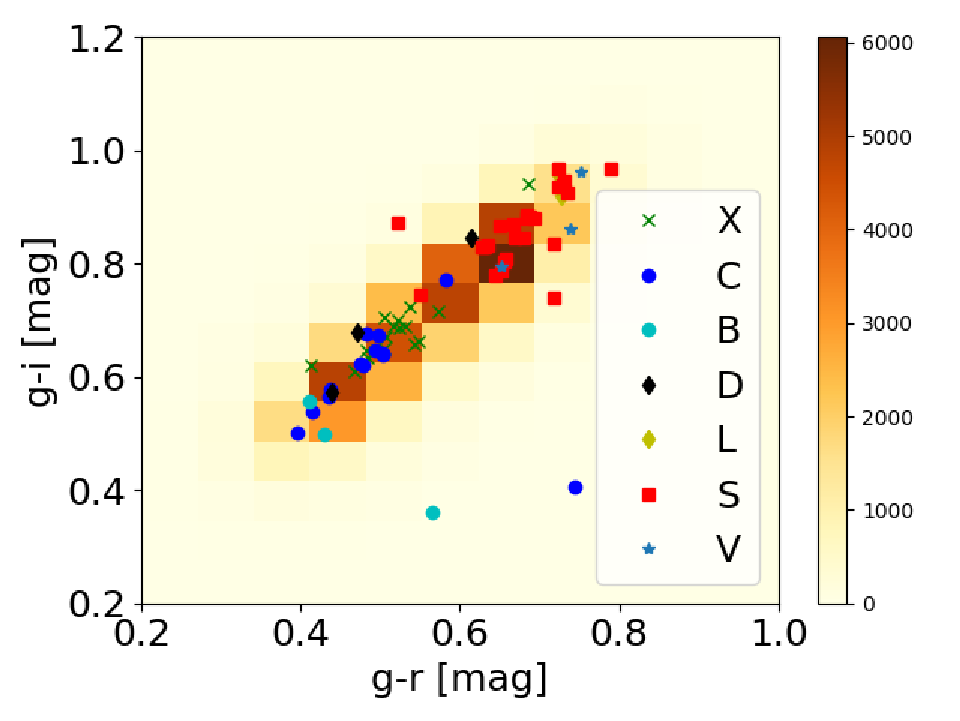}
  \caption{A 2D histogram in the $(g-r,g-i)$ plane of the asteroids with
    DES data.  The color symbols are objects for which taxonomical information
    is available in the \citet{2002Icar..158..146B},
      \citet{2004Icar..172..179L}, and \citet{2009Icar..202..160D} surveys,
    according to the color and size code shown in the legend.}
\label{Fig: Taxonomy_DES}
\end{figure}

First, we investigate where asteroids in this new data set with
known taxonomies could be located. Our results
are shown in figure~\ref{Fig: Taxonomy_DES}
We identified 71 asteroids with taxonomical data:
4 B-, 14 C-, 3 D-, 15 X-, 29 S-, 2 L-, and 3 V-types.  B, C, D, and X
asteroids are darker objects, associated with the so-called C-complex,
while the other types are brighter objects associated with the S-complex.
Our data show that the two complexes are fairly separated in this
domain, with C-complex asteroids located in the left part of the ellipsoidal
distribution of $(g-r,g-i)$ values, and S-complex asteroids concentrating in
the right part. However, different taxonomies inside these complexes overlap
with each other.  While DES data could help discriminate between
C- and S-complexes, we do not have enough resolution in the $(g-r,g-i)$
domain to perform a more in-depth taxonomical analysis, including the V-type
asteroids.

\begin{figure}
  \centering
  \centering \includegraphics[width=3.0in]{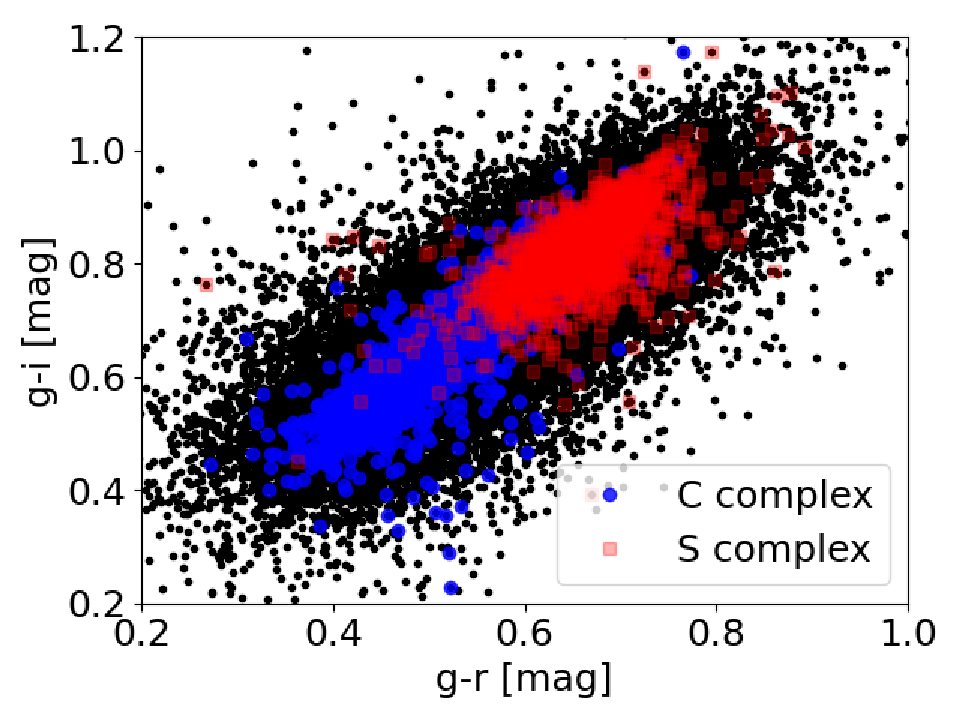}
  \caption{Projection in the $(g-r,g-i)$ plane of the asteroids with
    DES data and SDSS-MOC4 taxonomies. Blue circles identify asteroids likely
    to belong to the C-complex, while red squares display asteroids with
    more likely to belong to the S-complex. Black points are asteroids
    with no SDSS-MOC4 data.}
\label{Fig: SDSS_DES}
\end{figure}

To further confirm this hypothesis, we use the method of
\citet{2013Icar..226..723D} to obtain taxonomical information
for asteroids listed in the Sloan Digital Sky Survey-Moving Object Catalog data \citep[SDSS-MOC4,][]{2001AJ....122.2749I}.   Results for the
\citet{2010A&A...510A..43C}  and \citet{2018A&A...617A..12P} data sets
  are similar and will not be presented or discussed for the sake of brevity.
We identify 3347 asteroids
having data in both databases, of which 1514 are within the C-complex,
and 1833 are within the S-complex.
Figure~\ref{Fig: SDSS_DES} shows the $(g-r,g-i)$ distribution
of these objects, which confirms the results from the spectroscopical
surveys data: there is a clear separation between C-complex and S-complex
asteroids in this domain.  Again, the data resolution is
not sufficient to distinguish between different asteroid types, as previously
observed, and this is not shown in figure~\ref{Fig: SDSS_DES} for simplicity.

\begin{figure}
  \centering
  \centering \includegraphics[width=3.0in]{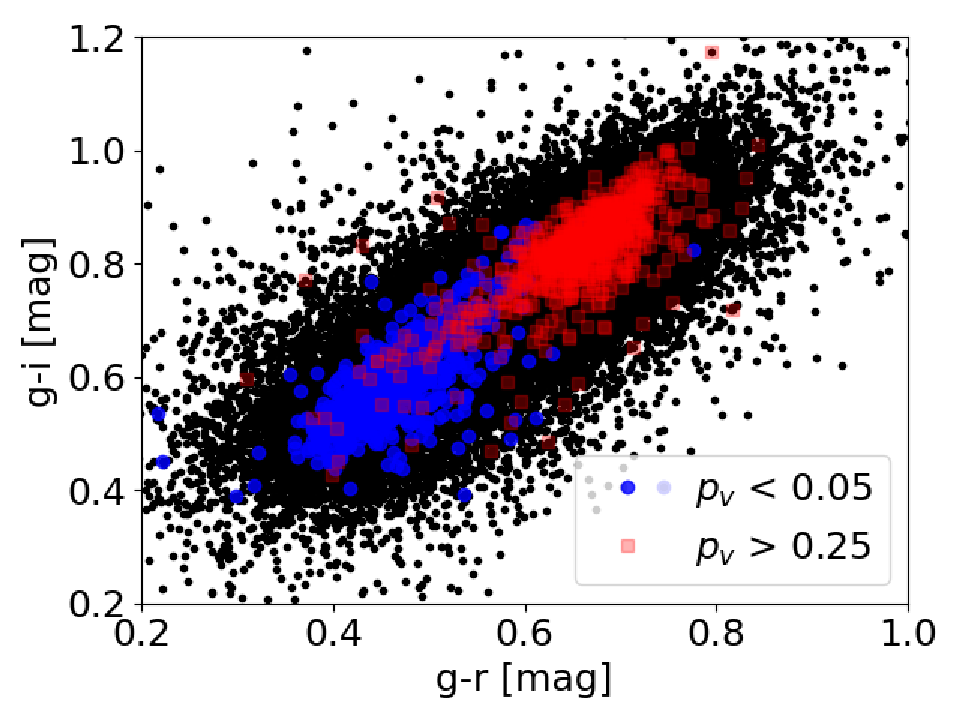}
  \caption{Projection in the $(g-r,g-i)$ plane of the asteroids with
    DES data and albedo values. Blue circles identify asteroids with
    albedo $p_V$ lower than 0.05, while red squares display asteroids with
    $p_V > 0.25$.}
\label{Fig: WISE_DES}
\end{figure}

Finally, while the distribution of albedo values may vary among a single
spectral type, C-complex asteroids tend to have lower values of geometric
albedos $p_V$, while S-complex ones have larger $p_V$.  We searched for
objects in the DES dataset with albedo data in the WISE and NEOWISE,
AKARI, or IRAS databases,
and identified 5122 asteroids, as shown in figure~\ref{Fig: WISE_DES}.
While asteroids with intermediate albedos ($0.05 < p_V < 0.25$) cover
the $(g-r,g-i)$ domain more or less uniformly, very dark ($p_V < 0.05$)
and very bright ($p_V > 0.25$) asteroids are 
quite separated in the $(g-r,g-i)$ domain, which confirms the trends
observed for taxonomic data.

Using taxonomic information obtained from the SDSS analysis, we utilized
genetic algorithms to determine the optimal ML methods for predicting the
complex type of asteroids based on the $(g-r,g-i)$ DES data. Our analysis
found that the Linear Support Vector Classifier (Linear SVC,
\citep{cortes1995support}) performed the best. 

Linear SVC is a type of
machine learning algorithm used for classification tasks, where the goal
is to assign each input to one of a set of predefined categories or classes.
Linear SVC is a variant of the Support Vector Machine (SVM) algorithm that
uses a linear kernel function. In Linear SVC, the algorithm tries to find the
hyperplane that separates the different classes with the largest possible
margin. The hyperplane is defined as the set of points in the feature space
where the decision boundary between classes lies. The margin is the distance
between the hyperplane and the closest points from each class. The larger the
margin, the more robust the classification will be to noise and outliers.

For our model, we used a $C$ parameter, which controls the trade-off
between achieving a low training error and a low testing error, of 0.1.
a penalty parameters equal to $l1$, which uses the $l1$ regularization
method, and a tolerance parameter, which
specifies the tolerance for stopping the optimization algorithm, of 0.0001.
Other parameters where the standard choices for the Linear SVC algorithm.
Our model achieved an accurarcy of 90.9\% on the validation set, which,
as in the previous section, was 20\% of the original training set.

\begin{figure}
  \centering
  \centering \includegraphics[width=3.0in]{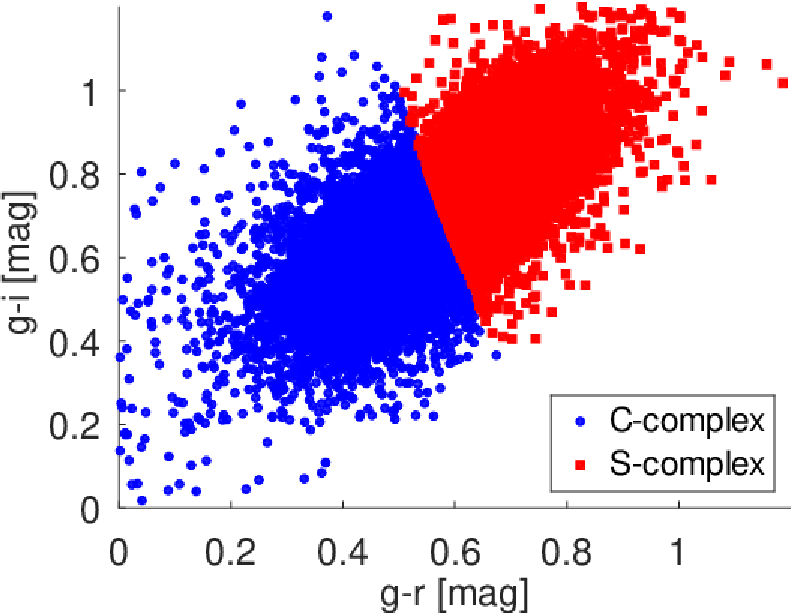}
  \caption{($g-r, g-i$) colors of the ML
    predicted data for asteroids in the DES database.}
\label{Fig: DES_gr_gi_gen}
\end{figure}

Figure~(\ref{Fig: DES_gr_gi_gen}) shows predictions for 58118 new
asteroids with no prior complex information. Our analysis suggests that 28871
of these objects belong to the C-complex, while 29247 are more likely to be
S-complex asteroids.  The higher fraction of S-complex asteroids, as
opposed to C-complex in the $(g-r, g-i)$ domain with respect to the
($gri$ slope,$i-z$) plane can be explained by the fact that, in the
$(g-r, g-i)$ domain V-type asteroids are considered as part of the
S-complex.

\section{V-type asteroids: DES candidates}
\label{sec: DES_V_type}

In section (\ref{sec: demeo_tax}), we observed that the accuracy of
predicting V-type objects in the $gri$ slopes and $i-z$ color plane was
higher (65.6\%) than for other classes. V-type objects are important
because of their association with differentiated parent bodies and basaltic
composition. Identifying their orbital location and physical properties may
provide insights about the early phases of our Solar System formation. To
date, asteroid 4 Vesta is the only confirmed differentiated body in the
main belt, but other possible differentiated parent bodies have been suggested
in the past. Using the DES data in the $gri$ slopes and $i-z$ color plane,
we identified 410 new potential V-type objects with available proper elements
from the Asteroid Family Portal AFP, http://asteroids.matf.bg.ac.rs/fam/
  \citep{2022CeMDA.134...34N}, 85 of which are located in the
central and outer main belt.

\begin{figure}
  \centering
  \centering \includegraphics[width=2.8in]{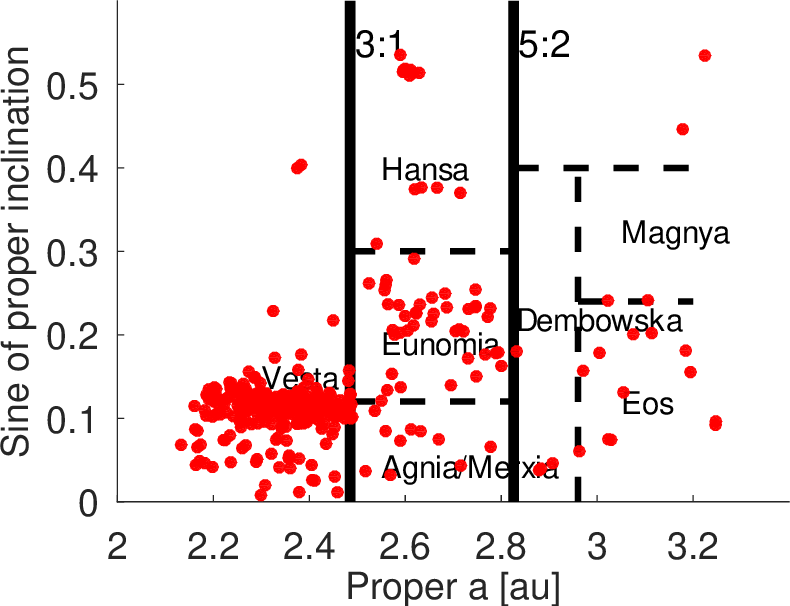}

  \caption{A proper $(a,\sin{(i)})$ projection of the V-type candidate
    asteroids identified using DES data. Vertical full lines identify the
    dynamical boundaries of the inner, central and outer main belt. The dashed
    lines show the location of the dynamical regions associated with possible
    local sources of V-type materials, as identified by
    \citet{10.1093/mnras/stu192} and \citet{2014MNRAS.444.2985H}.}
\label{Fig: DES_V_type}
\end{figure}

Figure~(\ref{Fig: DES_V_type}) displays their projection onto a proper
$(a,\sin{(i)})$ plane. The dynamical evolution of V-type objects suggests
that there may be six possible regions where injected material, either
from a local or remote source, can evolve due to non-gravitational forces.
\citet{10.1093/mnras/stu192} identified three such regions in the central
main belt, named after three possible local differentiated bodies: Hansa,
Eunomia, and the Agnya/Merxia parent body. A similar analysis by
\citet{2014MNRAS.444.2985H} found three comparable dynamical regions in the
outer main belt: the Dembowska, Eos, and Magnya regions. DES V-type
candidates are primarily located where they are expected, near the Vesta
family and in the densely populated Eunomia and Eos regions. Further
investigation of their physical properties and understanding their dynamical
evolution will remain a challenge for future studies.

\section{Summary and conclusions}
\label{sec: concl}

In this study, we aimed to explore the feasibility of using DES data
to infer physical properties of main belt asteroids. We initially focused on
using the $gri$ slopes and $i-z$ colors of asteroids, as this plane allows
us to employ the taxonomic classification scheme of
\citet{2013Icar..226..723D}. After removing outliers, we identified a
population of 17135 asteroids. However, with the exception of V-type and
possibly C-type asteroids, DES data in this plane is insufficient to
identify all the \citet{2013Icar..226..723D} taxonomies. Nonetheless, we can
still distinguish between C- and S-complex taxonomies. We utilized machine
learning approaches, optimized using genetic algorithms, to predict complex
labels for 10213 asteroids that previously had no taxonomic information.

A much larger sample of objects is available if we consider the
$(g-r,g-i)$ plane, with a sample of 61493 asteroids with such data obtained
after outlier removal. We employed machine learning algorithms to predict the
complex labels of 58118 new asteroids with no prior taxonomic information,
using the asteroids with SDSS-MOC4 known taxonomies as a training sample.

Lastly, we used DES data to identify 410 possible new V-type objects. Their
distribution in the proper $(a,\sin{(i)})$ domain is consistent with the
location of known V-type bodies. Future research could concentrate on
further examination of the physical characteristics and dynamical evolution
of the asteroids identified in this study.
  
\section*{Acknowledgments}

We are grateful to an anonymous reviewer for insightful and constructive
comments that greatly improved the quality of this work.
We would like to thank the Brazilian National Research Council
(CNPq, grant 304168/2021-1). J.I.B.C. acknowledges grants 305917/2019-6, 306691/2022-1 (CNPq) and 201.681/2019 (FAPERJ). This research used computational resources from the Interinstitucional Laboratory of e-Astronomy (LIneA) with financial support from the INCT of the e-Universo (Process number 465376/2014-2). F.S.F. acknowledges the support of Coordena\c{c}\~{a}o de Aperfei\c{c}oamento de Pessoal de N\'{i}vel Superior - Brasil (CAPES) - Finance Code 001.
V.F. acknowledges a CNPq support, PIBIC/ON (process 143944/2022-3). Funding for the DES Projects has been provided by the U.S. Department of Energy, the U.S. National Science Foundation, the Ministry of Science and Education of Spain, 
the Science and Technology Facilities Council of the United Kingdom, the Higher Education Funding Council for England, the National Center for Supercomputing 
Applications at the University of Illinois at Urbana-Champaign, the Kavli Institute of Cosmological Physics at the University of Chicago, 
the Center for Cosmology and Astro-Particle Physics at the Ohio State University,
the Mitchell Institute for Fundamental Physics and Astronomy at Texas A\&M University, Financiadora de Estudos e Projetos, 
Funda{\c c}{\~a}o Carlos Chagas Filho de Amparo {\`a} Pesquisa do Estado do Rio de Janeiro, Conselho Nacional de Desenvolvimento Cient{\'i}fico e Tecnol{\'o}gico and 
the Minist{\'e}rio da Ci{\^e}ncia, Tecnologia e Inova{\c c}{\~a}o, the Deutsche Forschungsgemeinschaft and the Collaborating Institutions in the Dark Energy Survey. 

The Collaborating Institutions are Argonne National Laboratory, the University of California at Santa Cruz, the University of Cambridge, Centro de Investigaciones Energ{\'e}ticas, 
Medioambientales y Tecnol{\'o}gicas-Madrid, the University of Chicago, University College London, the DES-Brazil Consortium, the University of Edinburgh, 
the Eidgen{\"o}ssische Technische Hochschule (ETH) Z{\"u}rich, 
Fermi National Accelerator Laboratory, the University of Illinois at Urbana-Champaign, the Institut de Ci{\`e}ncies de l'Espai (IEEC/CSIC), 
the Institut de F{\'i}sica d'Altes Energies, Lawrence Berkeley National Laboratory, the Ludwig-Maximilians Universit{\"a}t M{\"u}nchen and the associated Excellence Cluster Universe, 
the University of Michigan, NSF's NOIRLab, the University of Nottingham, The Ohio State University, the University of Pennsylvania, the University of Portsmouth, 
SLAC National Accelerator Laboratory, Stanford University, the University of Sussex, Texas A\&M University, and the OzDES Membership Consortium.

Based in part on observations at Cerro Tololo Inter-American Observatory at NSF's NOIRLab (NOIRLab Prop. ID 2012B-0001; PI: J. Frieman), which is managed by the Association of Universities for Research in Astronomy (AURA) under a cooperative agreement with the National Science Foundation.

The DES data management system is supported by the National Science Foundation under Grant Numbers AST-1138766 and AST-1536171.
The DES participants from Spanish institutions are partially supported by MICINN under grants ESP2017-89838, PGC2018-094773, PGC2018-102021, SEV-2016-0588, SEV-2016-0597, and MDM-2015-0509, some of which include ERDF funds from the European Union. IFAE is partially funded by the CERCA program of the Generalitat de Catalunya.
Research leading to these results has received funding from the European Research
Council under the European Union's Seventh Framework Program (FP7/2007-2013) including ERC grant agreements 240672, 291329, and 306478.
We  acknowledge support from the Brazilian Instituto Nacional de Ci\^encia
e Tecnologia (INCT) do e-Universo (CNPq grant 465376/2014-2).

This manuscript has been authored by Fermi Research Alliance, LLC under Contract No. DE-AC02-07CH11359 with the U.S. Department of Energy, Office of Science, Office of High Energy Physics.

\section{Data availability}

The taxonomical complexes data, as well as the data on DES V-type candidates,
are available at the NASA Planetary Data System (PDS): Carruba, V., Camargo, J. I. B., Aljbaae, S, and the Dark Energy Survey Team (2024). Taxonomy, colors, and slope parameters for asteroids from the Dark Energy Survey V1.0. urn:nasa:pds:gbo.ast.des.taxonomy::1.0. NASA Planetary Data System; https://doi.org/10.26033/m95p-bn08.

\section{Code availability}

All codes are available from the authors, upon reasonable request.

\bibliographystyle{mnras}
\bibliography{mybib}

\bsp

\label{lastpage}

\end{document}